\documentclass[useAMS,usenatbib]{mn2e}
\usepackage{graphicx}
\usepackage{natbib}  
\usepackage{epstopdf}
\newcommand\aj{{AJ}}%
\newcommand\araa{{ARA\&A}}%
\newcommand\apj{{ApJ}}%
\newcommand\apjl{{ApJ}}%
\newcommand\apjs{{ApJS}}%
\newcommand\aap{{A\&A}}%
\newcommand\aaps{{A\&AS}}%
\newcommand\mnras{{MNRAS}}%
\newcommand\pasp{{PASP}}%
\def\simgt{\lower.5ex\hbox{$\; \buildrel > \over \sim \;$}}
\def\simlt{\lower.5ex\hbox{$\; \buildrel < \over \sim \;$}}

\def\vmi{\hbox{\it V--I\/}}                                 
\def\bmi{\hbox{\it B--I\/}}                                 

\title[Relative ages of the Globular Clusters NGC~1851 and M4]{The influence of the C+N+O abundances on the determination of the relative
ages of Globular Clusters: the case of NGC~1851 and NGC~6121 (M4)   \thanks{
Based in part on observations made with the European Southern Observatory
telescopes obtained from the ESO/ST-ECF Science Archive Facility. }
\thanks{ This paper makes use of data obtained from the Isaac Newton Group Archive
which is maintained as part of the CASU Astronomical Data Centre at the
Institute of Astronomy, Cambridge.}}
\author[F. D'Antona, Stetson, Ventura, Milone, Piotto \& Caloi]
{F. D'Antona$^{1,2}$, 
P. \ B. Stetson$^{3}$,
P. Ventura$^{1}$, A. \ P. Milone${^4}$,
 G. Piotto${^4}$, V.Caloi$^{2}$ \thanks{E-mail: ventura, dantona @oa-roma.inaf.it;
 peter.stetson @nrc-cnrc.gc.ca; vittoria.caloi @iasf-roma.inaf.it; antonino.milone, giampaolo.piotto @unipd.it}\\
$^{1}$ INAF, Osservatorio Astronomico di Roma, Via Frascati 33, 00040 Monteporzio Catone (Roma), Italy.\\
$^{2}$ INAF, IASF--Roma, via Fosso del Cavaliere 100, I-00133 Roma, Italy\\
$^{3}$ Dominion Astrophysical Observatory, Herzberg Institute of
Astrophysics, National Research Council, 5071 West Saanich Road, \\ 
Victoria, British Columbia V9E~2E7, Canada\\
$^{4}$ Dipartimento di Astronomia, Universit\`a di Padova, 
Vicolo dell' Osservatorio 3, Padova, I-35122, Italy
}
\begin{document}
\date{Accepted . Received ; in original form }

\pagerange{\pageref{firstpage}--\pageref{lastpage}} \pubyear{2006}

\maketitle

\label{firstpage}

\begin{abstract}

The color magnitude diagram (CMD) of NGC~1851 presents two subgiant branches (SGB), probably due the presence of two
populations differing in total CNO content. 
We test the idea that a difference in total CNO may simulate an age difference 
when comparing the CMD of clusters to derive relative ages. We compare NGC~1851 with 
NGC~6121 (M4), a cluster of very similar [Fe/H]. 
We find that, with a suitable shift of the CMDs that brings the two
red horizontal branches at the same magnitude level, the unevolved main sequence and red giant branch 
match, but the SGB of NGC~6121 and its red giant branch ``bump" are {\it fainter} than
in NGC~1851. In particular, the SGB of NGC~6121 is even slightly fainter than the the faint SGB
in NGC~1851. Both these features can be explained if the total
CNO in NGC~6121 is larger than that in NGC~1851, even if the two clusters are coeval. 
We conclude by warning that different initial 
C+N+O abundances between two clusters, otherwise similar in metallicity and age,
may lead to differences in the turnoff morphology that can be easily attributed
to an age difference.
\end{abstract}


\section{Introduction}
\label{sec:intro}
The ages of Globular Clusters (GCs), the oldest systems populating our Galaxy
for which an age estimate is feasible, 
provide a lower limit to the age of the Universe, and therefore constitute one of
the important subjects of stellar astrophysical research.
Unfortunately, absolute age measurements are still affected by
a large number of uncertainties, in particular due to significant
uncertainties in GC distances and reddening values \citep[e.g.][]{gratton2003}.
Attention is then mostly devoted to the determination of
relative ages. These can be obtained with much higher precision, by measuring
the position of the main-sequence (MS) turnoff (TO) —--which
is the most-used age indicator—-- relative to some other feature in
the color--magnitude diagram (CMD), whose location has little or
no dependence on age, like the horizontal branch (HB) \citep{stetson1996,sarajedini1997}. Relative
ages provide fundamental information on the formation and early evolution of the Galaxy.\\
Several procedures, and better and better data sets have been employed to 
derive relative ages \citep{buonanno1998,salarisweiss1998,rosenberg1999,VandenBerg2000,
deangeli2005,marinfranch2009}.
The latest work by \cite{marinfranch2009} employs the homogeneous photometric
database achieved for 64 GCs in the ACS Survey of Galactic GCs
\citep[the HST Treasury program][]{sarajedini2007}), and compares the relative positions
of the clusters' MS TOs, using MS fitting to 
cross-compare clusters within the sample. According to the authors, the 
method provides relative ages to a formal precision of 2\% -- 7\%. 
One of the problems of the Mar\'in-Franch et al. study is that it does not include any 
consideration of possible systematic elemental abundance differences among GCs
that apparently have the same [Fe/H]. 
This problem is clearly to be attacked, now that we know that GCs 
may harbour different stellar populations, whether 
because of differences in the helium content \citep{dantonacaloi2004,
norris2004,dantona2005,piotto2005,piotto2007}; differences in $\alpha$-element
abundances in general \citep{ivans1999}; 
or differences specifically in total C+N+O, as will be discussed below. 
Limited helium content differences (in mass fraction, $\delta Y \sim 0.04-0.05$) 
do not seem to affect the TO luminosity at a given age \citep{dantona2002}, 
and so in principle do not affect the 
relative age determination. In contrast, stellar modelling shows that 
an increase in the total C+N+O content shifts the
isochrone turnoffs to lower luminosity \citep{bazzano1982, cassisi2008, ventura2009a},
without significantly changing the position of the unevolved MS, giant branch or HB.

Recently we have obtained observational evidence of the influence of 
such an additional parameter. In fact, two well separated subgiant branches (SGB)
are present in the CMD of the cluster NGC~1851 \citep{milone2008}. 
The faint SGB may be interpreted as being populated by stars either
older by $\sim$1Gyr, or having equal (or slightly younger) age 
and an increase in the total C+N+O content \citep{cassisi2008, ventura2009a}.
Formation models for multiple populations in GCs are still very much in their
infancy \citep{bekki-norris2006,bekki2007,dercole2008}, 
but no current model can easily accomodate a second
stellar generation younger by 1~Gyr. On the other hand, a total C+N+O spread among the stars in NGC~1851
does have an independent observational basis \citep{yong2008,yong2009}. An interpretation 
based on a different total C+N+O content is a variant of the ``two stellar generations"
models for GC formation that have already been investigated for the case of multiple main sequences,
possibly showing differences in helium content. In this scheme, the second
stellar generation forms out of matter polluted by
the stellar winds of massive asymptotic giant branch stars of the first generation \citep{cottrell-dacosta,
ventura2001}, although many details of such a model still need to be worked out \citep{dercole2008}. 

\begin{figure}
\includegraphics[width=8.1cm]{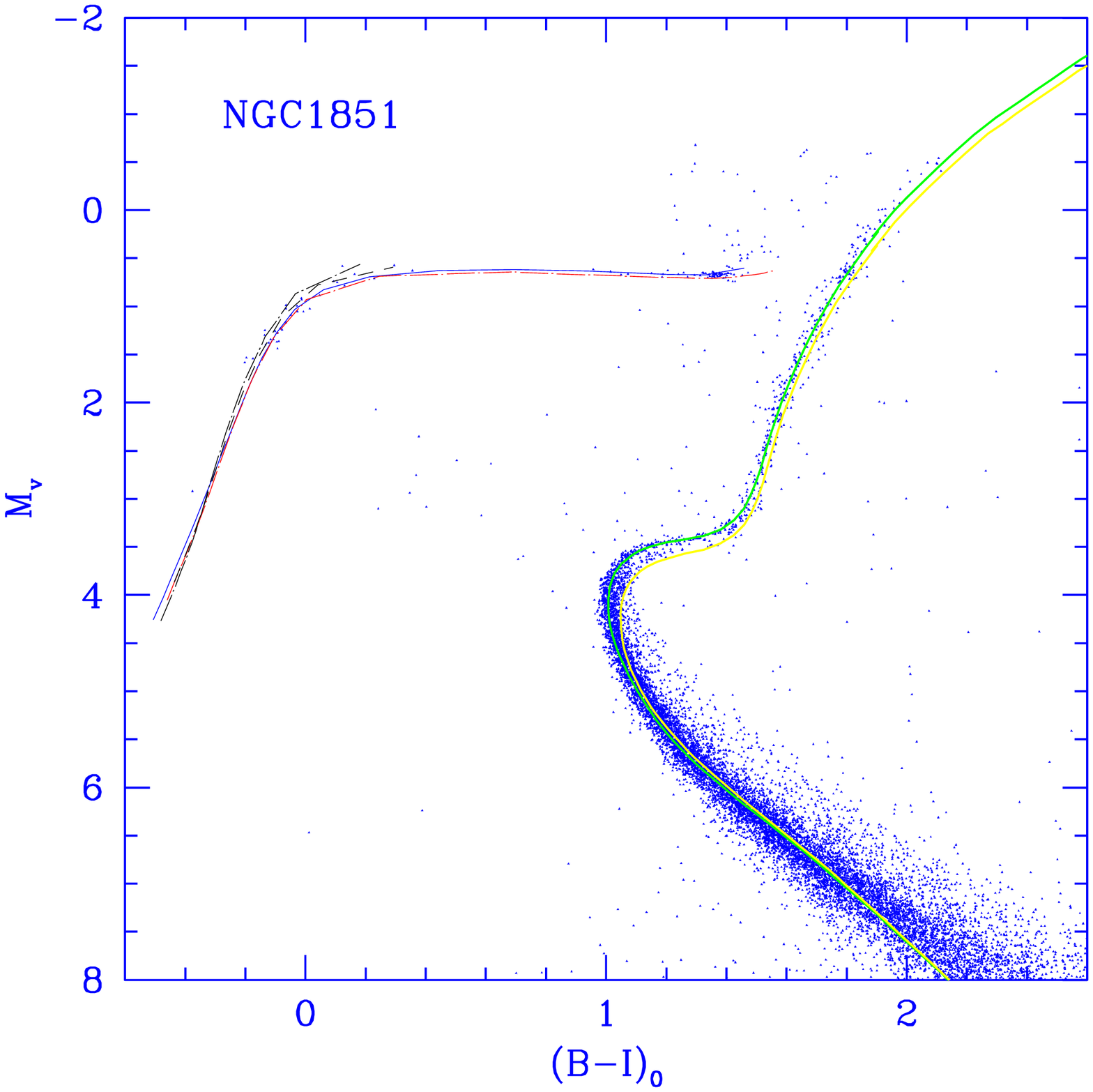}
\includegraphics[width=8.1cm]{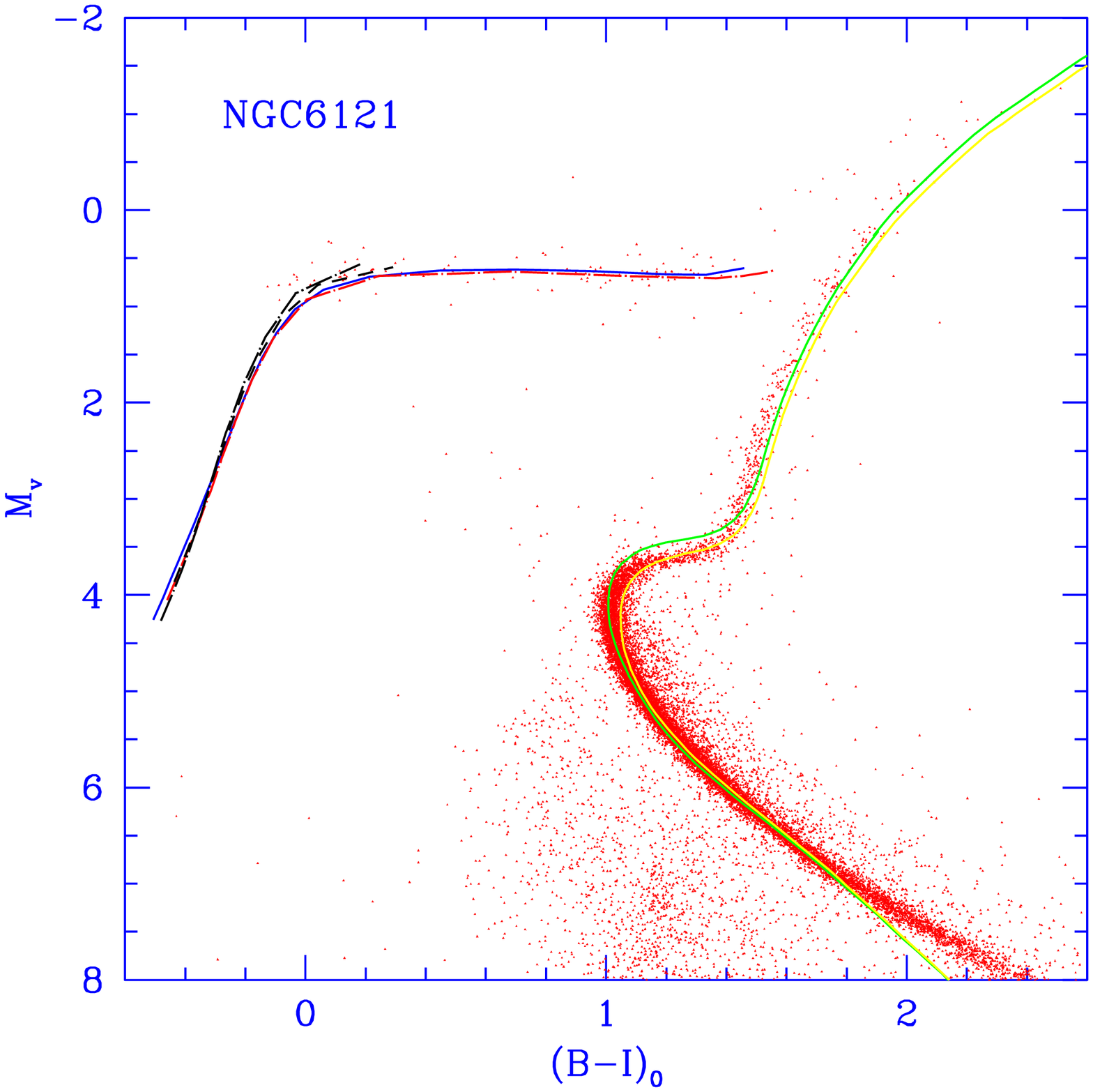}
\caption{Full view of the CMD diagrams in $V$ vs. \bmi\ of the clusters NGC~1851 (top figure) and M4
(bottom), superposed to the isochrones by applying magnitude and color shifts as 
explained in the text. For age t=12~Gyr and Y=0.24, the standard 
(CNO$\times$1)  Z=0.0010\ isochrone (green) and the isochrone CNO$\times$5 (Z=0.0035, yellow) are shown as full lines. 
The mixtures are described in Ventura et al. (2009a). 
The full (blue) line is the ZAHB locus of models by Ventura et al. (2009b) for Z=10$^{-3}$\ 
and Y=0.24, the dash--dotted (red) line is the (slightly dimmer) ZAHB locus 
for the Y=0.24, CNO$\times$5 mixture. The black lines on the blue side of the HB 
are the ZAHB locations for the mixture CNO$\times$3, Y=0.26 (dashed) and 
CNO$\times$5, Y=0.28 (dash--dotted).}
      \label{f1}%
\end{figure}
Now that we have such observational evidence, we can pose the question whether 
relative age indicators already present in the literature can be interpreted differently. In the following,
we explore the case of the relative ages of the two clusters NGC~1851 and NGC~6121 (M4), that
have always been considered twins from the point of view of metal abundance 
\citep[e.g.][]{zinnwest1984, rutledge1997,rosenberg1999,marinfranch2009}, with NGC~1851 somewhat 
younger if age is the only second parameter.  

The influence of varying the CNO abundances in stellar models is well documented in the astrophysical
literature (e.g. Simoda \& Iben 1970, Bazzano et al. 1982; see also \cite{vdbbell2001}
and references therein). In the last decade, however, the attention was mainly devoted to the effect of variations
in oxygen --and in the other $\alpha$--elements, following the discovery that population II stars
had a larger than solar [O/Fe] \citep[e.g.][]{mcwilliam1997}. 
An increase by 0.3~dex in [O/Fe] implies a 1~Gyr reduction in age for a fixed turnoff luminosity. 
Recently the attention has
been focused on the possible differences in total C+N+O deriving from the presence of multiple
stellar generations, so different mixtures of these elements have been considered. \cite{cassisi2008}
adopt an enhanced mixture based on a combination derived
from some observations of abundance anomalies, while \cite{ventura2009a} resorted more on the
chemical yields provided by theoretical massive AGB models. In any case, we are still far from a complete
investigation of the relative role of the three elements, that may become more 
necessary following the results of the present work. We base our considerations 
mainly on the models computed to explain the double subgiant branch in NGC~1851. 
This problem is very timely, as it is also confirmed by the
very recent exam by \cite{marin2009} of the influence, of CNO and helium 
variations, on the relative age determination of GCs, purely  from a theoretical point of view. 

\section{The two clusters NGC~1851 and NGC~6121}

\begin{figure}
\includegraphics[width=8.1cm]{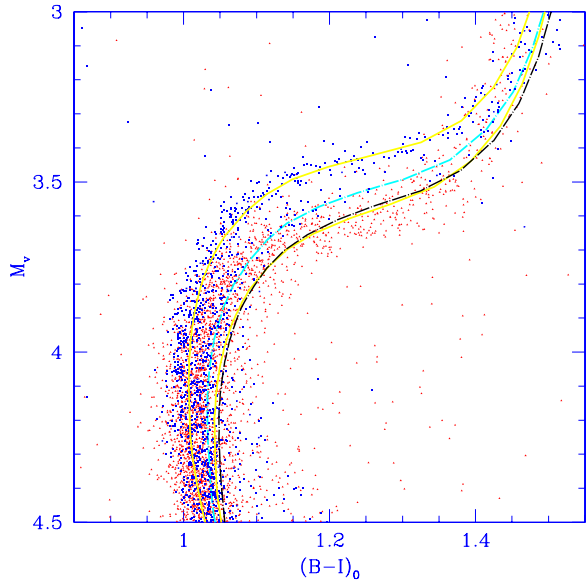}
\caption{Location of the TO and MS of the two clusters, for the magnitude and
color shifts adopted for Fig \ref{f1}. The standard (CNO$\times$1) isochrones for t=12 
and 14~Gyr with  Z=10$^{-3}$\ and Y=0.24 are shown as full (yellow) lines. 
The long-dashed (cyan) curve represents the 12~Gyr isochrone for models having 
Z=10$^{-3}$\ and CNO$\times$3, which fits the faint SGB of NGC~1851 well,
while the dot-dashed (black) curve is the 12~Gyr isochrone for CNO$\times$5.
}
      \label{f2}%
\end{figure}

In this analysis we exploit the archive of homogeneous CCD photometry maintained
by one of us \citep[see, e.g.,][]{stetson1998, stetson2000, stetson2005}.  The
photometry of NGC~1851 has already been described in \cite{milone2009}; their
CMD is reproduced here as the blue points in Fig.~1.  Note that the ground-based photometry
confirms the split of the SGB into faint and bright branches, as originally
found from HST data \citep{milone2008}. In addition, the fit by
\cite{ventura2009a} of the bright SGB with a standard isochrone of 12~Gyr, and
the fit of the corresponding faint SGB with the same isochrone and C+N+O
abundances increased by a factor three (CNO$\times$3 mixture) is consistent with the observations in
the $V$ vs. \bmi\ plane (see Fig.\ref{f2}).  

The photometry of NGC~6121 to be discussed here is similarly obtained from
analysis of 1,199 individual CCD images resulting from 814 exposures (55 of
the exposures had been taken with the eight-CCD WFI camera on the ESO/MPE 2.2m
telescope) made during ten observing seasons on six telescopes (ESO/Dutch 0.9m,
CTIO 0.9m, Jacobus Kapteyn 1m, ESO/Danish 1.5m, and Nordic Optical 2.6m Telescope, 
in addition to the aforementioned ESO/MPE 2.2m).  Any given star may have been
measured in as many as 255 $B$-, 202 $V$-, and 253 $I$-band images.  The results
have been transformed to the photometric system of Landolt (1992) following
methodology described in \cite{stetson1998, stetson2000, stetson2005}.

Determination of relative ages requires complex procedures for determining observational
parameters and for comparison with theoretical isochrones. Let us look at the final comparison
between the two clusters under discussion in the analysis of \cite{deangeli2005} and \cite{marinfranch2009}
summarized in Table~1. Both works make a very careful analysis, whose 
final results do not depend substantially on the adopted abundance scale and isochrones. 
Here we report the result for the \cite{carretta-gratton} abundances,
and for the models by \cite{cassisi2004} in De Angeli, and by \cite{dotter2007} in Mar\'in-Franch. 

\begin{table}
\caption{Relative ages for NGC~1851 and NGC~6121}             
\label{ages}      
\centering          
\begin{tabular}{c c c c c c c c c}     
\hline\hline       
Name & [Fe/H]$_{CG}$ & relative ages  & relative ages   \\ 
     &               & De Angeli 2005 & Mar\'in Franch 2009  \\ 
\hline            
NGC~1851 & --1.03 &  0.82 $\pm$ 0.07 &  0.78 $\pm$ 0.04     \\
NGC~6121 & --1.05 &  0.91 $\pm$ 0.04 &  0.98 $\pm$ 0.05     \\
\hline 
\hline
\end{tabular}
\end{table}

The table shows that---according to the age interpretation of the observed
differences---NGC~1851 is $\sim$10\% younger than NGC~6121 in the
\cite{deangeli2005} analysis, and $\sim$20\% younger in the analysis by
\cite{marinfranch2009}.  Assuming an age of 12~Gyr for NGC~6121,
NGC~1851 turns out to be a bit less than 11~Gyr (De Angeli), or ~9.5~Gyr
(Mar\'in-Franch) old, an age difference that is significant for models of
the formation of the Galaxy.  
\begin{figure}
\includegraphics[width=8cm]{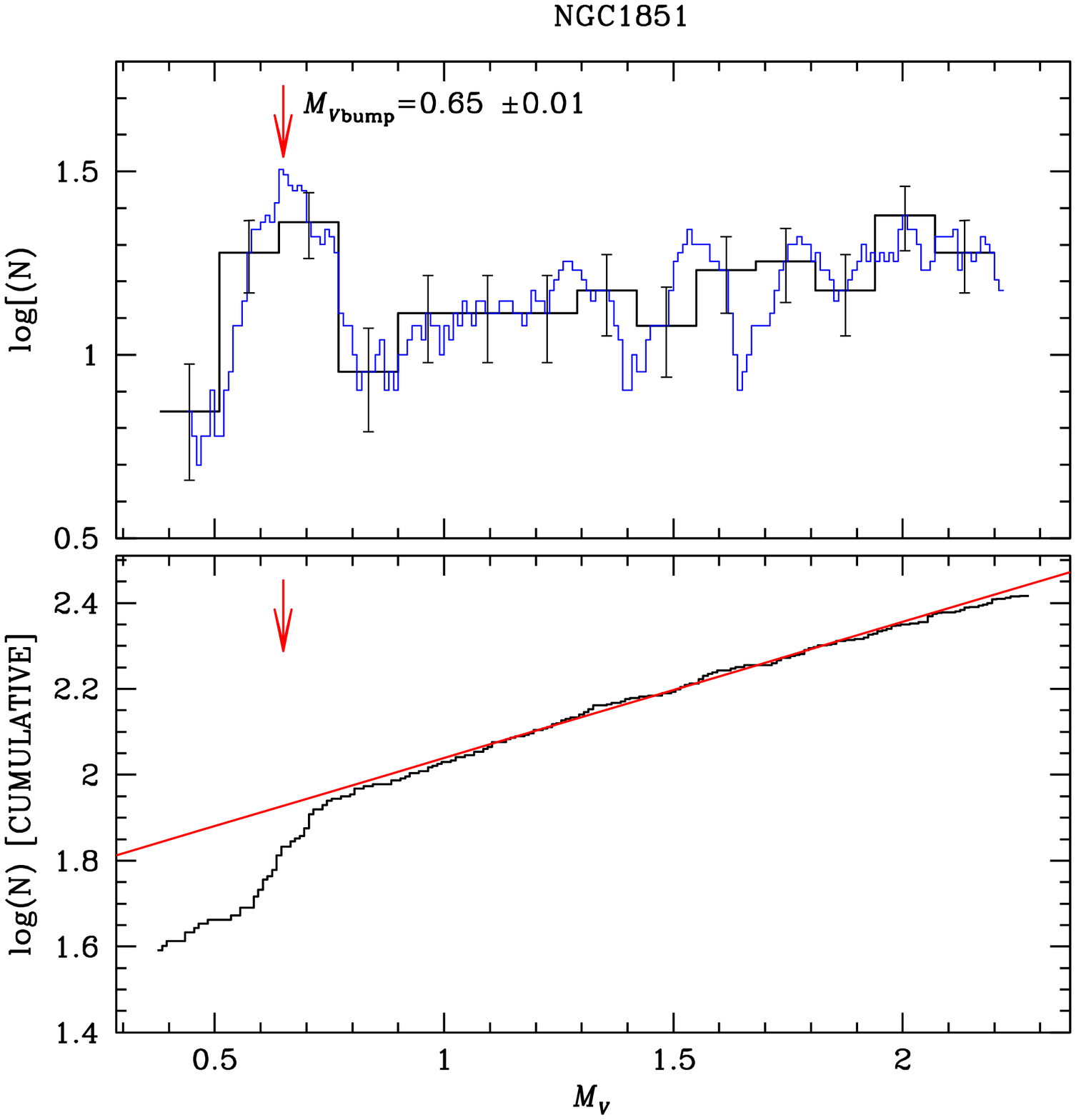}
\includegraphics[width=8cm]{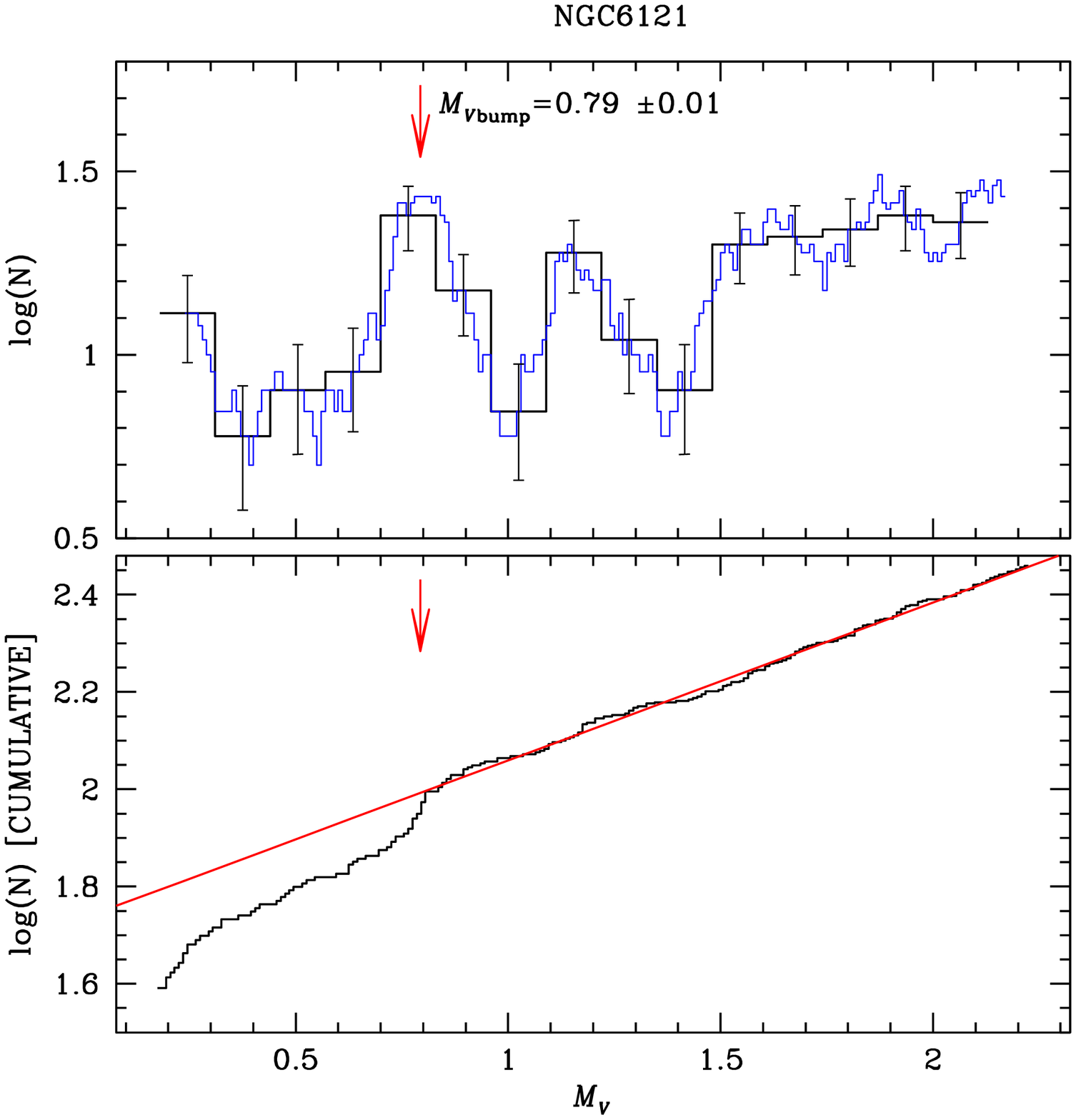}
\caption{Differential (top panels in each figure) 
and cumulative luminosity functions of the RGBs in NGC~1851 and NGC~6121.
For each cluster, we plot two histograms. The black ones are calculated by counting the red giants 
into intervals of 0.13\,mag in ${\it V}$.
Blue histograms are constructed along the RGB with a step of 0.013
magnitudes (a tenth of the bin size) in ${\it V}$, by
shifting the bin of fixed width of 0.13\,mag, with successive steps of
0.013 magnitude. The (red) arrows mark the location of the bump. 
For NGC~6121 the bump is clearly identified in both the differential and cumulative luminosity
functions, while in NGC~1851 the bump feature is broader in magnitude.  
This more complex bump structure may indeed reflect the
fact that NGC~1851 itself contains two populations with different C+N+O abundance.
}
      \label{f3}%
\end{figure}

In their Figure 7, \cite{deangeli2005} compare the $V$ {\it vs.} \vmi\ CMDs of these two clusters,
showing that, when the turnoffs are superposed, the HB of NGC~6121
turns out more luminous than the HB of NGC~1851, indicating 
indeed that NGC~6121 must be older. We make the reverse procedure. We use the
metallicity information, to superpose the red parts of the HB: even in the presence
of multiple populations, the red HB clumps are supposed to reflect the composition of the
first stellar generation, so that their location is presumed to be unaffected, e.g., by altered
helium. 
The match is made in the $V$\ versus  \bmi\ plane and is shown in
Fig.~\ref{f1} and in the enlarged view of Fig.\ref{f2}. 
We use for this comparison the absolute magnitude $M_{V}$, by positioning the
red clumps (RHB) of NGC~1851 and NGC~6121 on the zero-age horizontal branch (ZAHB) of our theoretical
models \citep{ventura2009b}, drawn in the figure. 
We selected by eye all the stars that, on the basis of their position in
the CMD, belong very likely to the RHB and calculated their median ${\it
V}$ magnitude. We find $V_{HB}$=16.19$\pm$0.01  and $V_{HB}$=13.44$\pm$0.01 for NGC 1851
and NGC 6121 respectively. The error associated to the HB position is calculated as
$\sigma/\sqrt(N-1)$ where N is the number of red HB stars and $\sigma$ is the $68.26^{th}$
percentile of the absolute deviation from the average of their $\it {V}$ distribution.
For the aims of this paper however we do not need the individual distance moduli of the
two clusters (e.g., we could have chosen smaller values, assuming a
smaller age) but their relative displacement, to which we can then associate 
an error of $\pm$0.01~mag.
We also choose color shifts so that the red giant branch (RGB) locations coincide. 
The apparent $V$\, magnitudes of NGC~1851 are shifted by $\delta V$=--15.52mag and the 
colors are shifted by $\delta$(\bmi)=--0.08. The corresponding shifts for NGC~6121 are
$\delta V$=--12.77mag and $\delta$(\bmi)=--1. 
On the blue side of the HB, the stars in M4 are slightly more luminous than in NGC~1851. 
This feature is not important for the present discussion, but remember 
that a slightly larger helium abundance
in the blue HB of M4 may lead to greater blue HB luminosity, as shown by the (blue) ZAHBs with Y=0.26 
(dashed) and Y=0.28 (dash--dotted) shown in Fig.\ref{f1}. We see in the figure
that the two MSs are well matched. The isochrones of
12 and 14~Gyr by \cite{ventura2009a} for Z=$10^{-3}$\ and standard C+N+O abundances
are also shown. The TO locations indicate
an age difference of $\sim$2~Gyr, as also concluded by \cite{marinfranch2009}
Since {\sl the SGB and TO of NGC~6121 appear fainter than the faint SGB of NGC~1851}, another
interpretation is possible: if NGC~6121 stars have total C+N+O abundance even larger than that of the
stars on the faint SGB of NGC~1851, they could be coeval with the stars in NGC~1851. This is illustrated 
in Fig~\ref{f2} by the 12~Gyr isochrone by \cite{ventura2009a} for a
composition in which the total CNO is enhanced by a factor of five (CNO$\times$5).


Are there any independent indications that this solution (same age and large 
difference in total C+N+O abundances) is to be preferred? A possible hint comes 
from Fig.~\ref{f3}, which shows the differential and
cumulative red giant luminosity functions, and the location of the red giant ``bump" 
due to the penetration of
the hydrogen burning shell into the mean molecular weight discontinuity (or $\mu$-barrier) 
left by the maximum extension in mass fraction
reached by the convective envelope at the beginning of giant-branch evolution. 
Standard theory and observations do not fully agree on the location in magnitude of the bump. 
Many models---including the \cite{ventura2009a} models---provide bumps
$\sim$0.25mag more luminous than the observed ones \citep{zp2000}. This may be due
to uncertainties either in the metallicity scale \citep[e.g.]{zoccali1999} or 
in the mixing below the convective envelope \citep{girardi2000}, or to additional
mixing caused by rotation \citep{palacios2006}.  
Whatever the causes of this small discrepancy, here we are
interested mainly in the relative positions of the bumps in the two clusters under discussion.  
Applying the distance moduli used to build up Fig.\ref{f1}, to which we can assign a relative error
not larger than $\pm$0.01~mag, we see that the absolute magnitudes (median value) of the RG bump are
M$_v$=0.65$\pm$~0.01\,mag 
and M$_v$=0.79$\pm$0.01\,mag, respectively for NGC~1851 and NGC~6121. 
To determine the bump locations, we selected by hand stars that, according to their
position in the CMD, have a high probability to belong to the RGB bump
and calculated their median ${\it M_{\rm V}}$ (${\it M_{\rm V}}_{\it
bump}$). The error is calculated as: $\sigma/\sqrt(N-1)$, where $\sigma$ is 
the $68.26^{th}$ percentile of the absolute deviation from the average of 
the ${\it M_{\rm V}}$ distribution and N is the number of selected stars.
We see that there is a difference of 0.14$\pm 0.02$\,mag in the bump level. 
Can we change the relative distance of the clusters in order to match the bumps?
The red HB luminosity is indeed the best standard parameter for stars having the same [Fe/H]. 
Even if we assume that the HB red clump is populated by stars with
similar [Fe/H] content but different C+N+O, the models by \cite{ventura2009b}
show that the ZAHB luminosity at the red side of the HB decreases only by $\sim 0.04$\,mag, for 
a five-fold increase in C+N+O, so the HB location can not explain this difference in the bump magnitude.
An age increase of $\sim$2\,Gyr can not explain the difference either: according to our models, 
the bump location becomes dimmer by only $\sim$0.05\,mag. 
On the other hand, the models by \cite{ventura2009a} for Z=10$^{-3}$
and different C+N+O contents show that the bump luminosity {\it decreases} (almost) linearly with
the C+N+O increase. It was already known that an increase in
the oxygen abundance produces a lower bump luminosity \citep[e.g.][]{vdbbell2001}. 
We obtain $\delta M_v$(bump)$\simeq 0.037$ times the C+N+O 
abundance increase with respect to the ``standard", $\alpha$--enhanced, composition (CNO$\times$1).
A difference of 0.14$\pm 0.02$\,mag, for a fixed age and helium abundance, 
corresponds to the difference between CNO$\times$1 and CNO$\times (3.8\pm 0.6)$\, bumps.
We conclude that the difference in absolute magnitude of the bumps is consistent
with the hypothesis that M4 stars have a C+N+O content a factor of $\sim$4 larger than the bright 
SGB stars of NGC~1851.

Note also that the bump peak is more dispersed in NGC~1851 than in NGC~6752.
This difference may indeed reflect the
fact that NGC~1851 itself contains two populations with different C+N+O abundances.

\section{Discussion: a plea for CNO abundance determinations and for a wider theoretical
exam}
The comparison in Figs. \ref{f1} and \ref{f2} has shown that the two clusters
we are considering may differ in age by about 2~Gyr 
\citep[as found by][]{marinfranch2009}. However, now that we know that NGC~1851
harbours two SGBs, and that the faint SGB is very probably 
made up of coeval stars richer in total
C+N+O than the bright SGB stars, it is tempting to propose that the stars
of NGC~6121 are also coeval with ``all" the stars of NGC~1851, but they have a global C+N+O
abundance larger even than the stars on the fainter SGB in the latter cluster.
An analysis of the magnitude difference of the RGB bump locations in the two clusters,
and the relative location of their SGBs are consistent with the hypothesis 
that the global C+N+O in M4 stars is about a factor $\sim$4
larger than in the main population of NGC~1851. 
At present, this conclusion must be taken more as a working hypothesis 
for further discussions 
than as a proven fact. We need both careful, extensive and homogeneous abundance determinations 
of the C+N+O abundances in most GCs, and further theoretical analysis of the role of the three different
elements---C, N and O---in shaping the cluster isochrones. Preliminary results of \cite{ventura2009a} 
suggest that the most important role is played by the late phases of hydrogen
burning close to the turnoff, where hydrogen burning shifts from p-p to the CN chain. As well
known, however, also the variations in oxygen, although they do not affect at 
all the turnoff evolution, where the ON cycle does not play a role, affect the 
opacities and the evolution by producing smaller ages \citep{vdbbell2001}.

From an observational point of view, it is not unlikely that detailed elemental abundances
may differ from cluster to cluster, even for the same iron abundance and global 
$\alpha$-element abundances. 
\cite{ivans1999,ivans2001}, in their study of a large sample of giants in M4 (NGC~6121) and
M5 (clusters with very similar metallicity), 
show that on average oxygen, silicon, aluminum, barium, and lanthanum 
are overabundant in M4 with respect to M5. The C+N+O 
 abundance sum is constant to within the observational errors. 
 These results are confirmed and enlarged by \cite{yong2008a}, who find that
silicon, copper, zinc, and all s-process elements are approximately 0.3 dex overabundant in M4
relative to M5. \cite{yong2008b} compared the rubidium and lead abundances
in M4 and M5, finding that M4 has global abundances of both
Rb and Pb much larger than M5, and that there is no star-to-star variation
in the abundances of these elements.  
As mentioned before, \cite{yong2009} find a large spread
in C+N+O among 4 giants of NGC~1851, confirming independently the interpretation of the
two SGBs in terms of different CNO abundances. In their Fig.4, they compare C+N+O abundances in NGC~1851
with those of the giants in M4 determined by \cite{smith2005}. These latter abundances show
a little spread, and fall in the middle of the distribution of NGC~1851 giants. Although this
result goes in the direction we suggest, the location
of M4's SGB relative to the NGC~1851 SGBs would require the CNO abundances in
M4 to be similar to or greater than the {\it largest} abundances of NGC~1851. 
The small sample examined, and the possible systematic differences
in the abundance analysis [\cite{yong2009} versus \cite{smith2005}] do not allow us
to derive
more stringent conclusions.

In summary, the opening of Pandora's box of the peculiar morphologies 
of so many CMDs of GCs may reopen the problem of relative age determination.
We should carefully examine the abundances of C+N+O in the clusters, especially those
showing different ages at similar metallicities, before the derived age spreads may be
attributed to the modalities of formation of the Galaxy.

\section{Acknowledgments} 
This work has been supported through PRIN MIUR 2007 
``Multiple stellar populations in globular clusters: census, characterization and
origin" (prot. n. 20075TP5K9).  PBS is pleased to thank H.~Bond, F.~Grundahl, A.~Rosenberg,
M.~Zoccali, and the Isaac Newton Group and ESO archives for contributing much of the data
employed here.

\label{lastpage}

\end{document}